\documentstyle[12pt]{article}
\begin{document}
{\rm

\title{Correlation and Finite Interaction-Range Effects  \\ in \\
High-Energy Electron Inclusive Scattering}

\author{Akihisa KOHAMA 
\thanks{Special Postdoctoral Researcher at RIKEN. 
e-mail:~ kohama@rikaxp.riken.go.jp}\\
RIBF Project Office, Cyclotron Center, \\
RIKEN (the Institute of Physical and Chemical Research), \\
2-1 Hirosawa, Wako-shi, Saitama 351-0198, JAPAN, \\
Koichi YAZAKI 
\thanks{Present address:~Physics Lab. Tokyo Woman's Christian
University, Suginami-ku, Tokyo 167-8585, JAPAN
}\\
Department of Physics, School of Science, \\
University of Tokyo, \\
7-3-1 Hongo Bunkyo-ku Tokyo 113-0033, JAPAN \\
and \\
Ryoichi SEKI \\
Department of Physics and Astronomy, \\
California State University, Northridge, CA 91330, and \\
W. K. Kellogg Radiation Laboratory, \\
California Institute of Technology, Pasadena,
CA 91125, USA }
     
\maketitle

\begin {abstract}
We calculate cross sections 
of high energy electron inclusive scattering 
off nuclear matter in a new and consistent formulation 
based on the Green's function method with the Glauber approximation, 
which is an extension of our previous work 
on the nuclear transparency in $(e, e'p)$ reaction. 
The comparison with other approaches is discussed. 
In this framework, 
we study the finite-range effect of the nucleon-nucleon interaction 
in the final-state interactions, 
and the effect of the nuclear short-range correlation. 
We propose a zero-range approximation, 
which works well when correlation and finite interaction-range 
effects are included. 
It greatly reduces the numerical work, 
while maintaining a reasonable accuracy. 
\end{abstract}

\newpage
\setcounter{section}{0}
\setcounter{equation}{0}
\section{Introduction}
\label{sec-intro}

Experimental data on high-energy electron-nucleus inclusive scattering 
have been accumulated for a decade \cite{Day:PRC}. 
The experiments were mostly done at SLAC, 
and will be done at the Jefferson Lab. 
Also a new facility 
called MUSES (Multi-Use Experimental Storage ring)
is planned at RIKEN \cite{MUSES}.
Precise data will become available with these facilities and 
provide us with a way of studying where and how 
the quark-gluon degrees of freedom may come into sight. 

The process that we study is 
the $(e, e')$ inclusive reaction (Fig.~1)
in the nucleon quasi-elastic region, {\it i.e.}, 
\begin{equation}
   e + A \rightarrow e' + N + (A - 1)^{\ast}. 
\end{equation}
The incident electron has a four-momentum, 
$k_{\mu}$ $= (\epsilon, {\bf k}_{\rm e})$, 
and the outgoing electron, 
$k'_{\mu}$ $= (\epsilon', {\bf k'}_{\rm e})$. 
The momentum transfer is $q_{\mu}$ $= k_{\mu} - k'_{\mu}$
$= (\omega, {\bf q})$. 
We are interested in the quasi-elastic region, 
{\it i.e.}, $| {\bf q} | \sim 2 \;[{\rm GeV/c}]$, and
$\omega \le 2 \;[{\rm GeV}]$. 
So the elementary process is considered to be 
mostly the electron-nucleon elastic scattering, 
$e + N$ $\rightarrow e + N$, 
and the final-state interaction (FSI) of the struck nucleon 
is what we wish to study here. 

In this work we calculate the effect of the FSI, 
taking into account only the nucleonic degrees of freedom. 
We use the Glauber approximation for the FSI, 
including the nuclear correlation. 
We believe that our treatment is a more systematic approach 
than the previous works. 

The theoretical treatment of this reaction is similar to 
that we used for the nuclear transparency in the $(e, e'p)$ reaction 
at large-momentum transfers \cite{Koh:NP1}, \cite{Koh:NP2}. 
This process has recently attracted  
many researchers in connection with a speculated phenomenon 
of the color transparency \cite{Bro:Pro}. 
The present formulation allows us to 
describe the $(e, e')$ and $(e, e'p)$ reactions in a unified way. 

The nuclear transparency for electron scattering can be defined as the 
ratio of the $(e, e'p)$ response to the inclusive $(e, e')$ response 
in the quasi-elastic region, 
and is conveniently formulated by the Green's function method \cite{MY:PP}. 
We will give here a way of calculating the response functions 
based on the Glauber approximation \cite{Glau:Lec} 
for the Green's function, without taking account of 
the proton internal structure. 

The contents of this paper are as follows:
In sec.~\ref{sec-form} we explain our formulation. 
We explain our standpoint about the elementary process, 
{\it i.e.}, the electron-nucleon cross section, in subsec.~\ref{sec-elem}. 
We derive a closed form of the inclusive cross section 
based on the Glauber approximation in subsec.~\ref{sec-glauber}. 
The definition of the FSI-function is also given there. 
The expression of the response function for nuclear matter 
is derived in subsec.~\ref{sec-expnm}, 
and the definition of our ``convolution" function is found. 
We derive the approximate expressions of the FSI-function 
when we apply zero-range approximations for the nucleon-nucleon potential 
in subsec.~\ref{sec-appr}. 
In sec.~\ref{sec-comp} we compare our treatment of the FSI with others. 
In subsec.~\ref{sec-grs} we discuss the difference and the relation 
between our formulation and the theory 
of Gersch, Rodriguez, and Smith \cite{GER:72}, \cite{GER:73}, \cite{RT:NPA}. 
A similar comparison with the optical potential 
formalism \cite{Ben:PRC}, \cite{Ben:PLB}
is given in subsec.~\ref{sec-opt}. 
In sec.~\ref{sec-nume} we show our numerical results 
and discuss their indications. 
We show the FSI-function in subsec.~\ref{sec-conv}, 
and the inclusive cross sections in subsec.~\ref{sec-xsec}. 
We summarize our results in sec.~\ref{sec-summary}. 
In the Appendix 
we show a method for deriving a nucleon-nucleon potential from 
the experimental scattering amplitude in the eikonal approximation.

\newpage
\setcounter{equation}{0}
\section{Formulation}
\label{sec-form}

In this section we explain our formulation 
of the electron-nucleus inclusive scattering based 
on the Green's function method \cite{MY:PP}. 
We employ the Glauber approximation for the FSI of the struck nucleon. 

\subsection{Choice of the $e N$ Cross Section}
\label{sec-elem}

Let us first explain our standpoint about 
the elementary process of the $(e, e')$ inclusive reaction. 

We start with the $(e, e')$ inclusive differential cross section 
on a target nucleus, $A$, 
which is assumed to be expressed as \cite{FW}
\begin{eqnarray}
  {d\sigma_{{\rm eA}} \over d\Omega d\omega}
  = \langle {d\sigma_{{\rm eN}} \over d\Omega} \rangle_{{\rm el, on}} \;
    S(\omega, {\bf q}),
\label{eq:form1}
\end{eqnarray}
where the cross section on the r.h.s.
is that of {\it on-shell} electron-nucleon elastic scattering 
for the same incident energy and scattering angle, 
and $\langle \ldots \rangle$ implies averaging over 
the spin, and the isospin, of the target nucleons. 
The difference between the longitudinal and the transverse responses 
is also neglected. 

This factorized form, eq.~(\ref{eq:form1}), is an approximation, 
on which some comments are in order: 
The nucleon struck by the electron is not in free space, 
but in the nuclear medium both in the initial and in the final states. 
Such a nucleon is often said to be {\it off-the-mass shell}, 
and its electromagnetic form factors to be used in the calculation 
of the cross section are generally different from those of 
the {\it on-shell} nucleon. 
However, we have no reliable way of estimating the differences. 
In fact, we need a model for the internal structure of the nucleon 
as well as a model of its interaction with the nuclear medium to 
study the off-shell effects. 
We note here that the off-shell effects are essentially dynamical 
and cannot be obtained by kinematical considerations 
\cite{defo:NP}, \cite{Nau:NP}. 
In the following we discuss how off-shell the nucleon will be 
in the case of inclusive $(e, e')$ on nuclear matter. 

Let us assume that the single nucleon spectrum in the nuclear matter 
is given by 
\begin{eqnarray}
  (E - V)^{2} = (m_{{\rm N}} + S)^2 + {\bf p}^2, 
\label{eq:elem1}
\end{eqnarray}
where $E$, ${\bf p}$ and $m_{{\rm N}}$ are the energy, the spatial 
momentum and the mass of the nucleon, respectively, 
with $V$ $(S)$ being the single particle vector (scalar) potential.
The degree of the nucleon being off-the-mass shell is measured by 
\begin{eqnarray}
  \delta m^{2} &\equiv& E^2 - {\bf p}^2 - m_{{\rm N}}^2 
  = 2 E V - V^2 + 2 m_{{\rm N}} S + S^2. 
\label{eq:elem2}
\end{eqnarray}
The Walecka model \cite{SW:Adv} gives 
$V$ $= 300\;[{\rm MeV}]$, $S$ $= -350\;[{\rm MeV}]$ (Case I). 
We will also consider the two extreme cases, {\it i.e.}, 
$V$ $= 0\;[{\rm MeV}]$, $S$ $= -50\;[{\rm MeV}]$ (Case II), and 
$V$ $= -50\;[{\rm MeV}]$, $S$ $= 0\;[{\rm MeV}]$ (Case III), 
keeping the sum, $S + V$, to be the same. 
Figure~	2 shows $\delta m^{2}/m_{{\rm N}}^2$ 
as a function of $E$ for these three cases. 
In all the cases, the nucleon in the initial state, 
where $E \simeq m_{{\rm N}}$, is not very off-shell, 
because $\delta m^{2}/m_{{\rm N}}^2$ $\simeq -0.1$. 
On the other hand, the nucleon in the final state, 
where $E$ $\simeq 2 - 3 m_{{\rm N}}$, is far off-shell, {\it i.e.}, 
$\delta m^{2}/m_{{\rm N}}^2$ $\simeq O(1)$, in the case (I), 
while it is not very off-shell in the cases (II) and (III). 
This exercise tells us that one should be careful in choosing 
the {\it off-shell} kinematics for the inclusive $(e, e')$ response. 
What may be important in the high-energy inclusive reaction is 
the off-shellness of the nucleon in the final state, 
not in the initial state. 
This has been stressed previously by,  
{\it e.g.}, Uchiyama, {\it et al.} \cite{Uchi:PLB},
and Ciofi and Simula \cite{Cio:PLB}. 

The off-shell cross section proposed so far is 
that of the half-off-shell \cite{defo:NP}, \cite{Nau:NP}. 
In the prescription the bound nucleon in the initial state is treated 
as off-the-mass shell, 
while the nucleon in the final state is treated on-the-mass shell. 
This is reasonable for the $(e, e'p)$ semi-inclusive reaction, 
but not for the $(e, e')$ inclusive. 
Furthermore, for the $(e, e')$ inclusive process 
the nuclear matter limit ($A \rightarrow \infty$)
can be taken, while for the $(e, e'p)$ semi-inclusive process 
that limit cannot be defined, because the nucleon in the final state 
cannot be free due to the infinitely extended nuclear matter. 
Therefore, the off-shell cross section, which is presently available, 
is not suitable for the elementary process of 
the $(e, e')$ inclusive reaction. 

We thus choose to avoid using such off-shell cross sections, 
and use the {\it on-shell} cross section. 
This is the basis for our choice of the {\it on-shell} kinematics 
for the initial nucleon. 
The construction of the models describing the internal structure of 
the nucleon and its interaction with the nuclear medium 
is beyond the scope of this work, 
and we use the simple factorized form with the on-shell form factors. 

Therefore the agreement of the numerical results with the experimental data 
is not necessarily the aim of this work. 
In the following we concentrate on a consistent treatment 
of the FSI of the struck nucleon in the Green's function method 
with the Glauber approximation.

\subsection{Green's Function Method}
\label{sec-glauber}

We now proceed to the derivation of the closed form 
of the response function, $S(\omega, {\bf q})$, in eq.~(\ref{eq:form1}) 
in the Green's function method with the Glauber approximation. 
The response function is defined by 
\begin{eqnarray}
  S(\omega, {\bf q})
  &\equiv& \sum_{X} |\langle X| \hat{O}({\bf q}) |A \rangle |^{2}\;
           \delta(\omega - E_{X}) 
\nonumber\\
  &=& -{1 \over \pi}\; {\rm Im}~  R(\omega, {\bf q}), 
\label{eq:form2}
\end{eqnarray}
where
\begin{eqnarray}
   R(\omega, {\bf q}) 
   \equiv \langle A | \hat{O}^{\dag}({\bf q}) \;
                     {1 \over \omega - \hat{H}_{A} + i \eta} \;
                      \hat{O}({\bf q}) \;
          | A \rangle,  ~~~(\eta > 0). 
\label{eq:form2a}
\end{eqnarray}
The initial state, $A$, is the ground state of the target nucleus, 
and the sum over the final states, $X$, is taken, 
because we are considering the inclusive process. 
$\hat{H}_A$ is the full hamiltonian of the target nucleus. 
The energy of the final state, $E_{X}$, in eq.~(\ref{eq:form2}) is 
measured from that of the ground state of the target. 

$\hat{O}({\bf q})$ in eq.~(\ref{eq:form2}) 
is a hard interaction operator defined by 
\begin{eqnarray}
  \hat{O}({\bf q})  &\equiv& 
  \exp \{{\it i} {\bf q}\cdot {\bf r}_{1}\}, 
\label{eq:form3}
\end{eqnarray}
where ${\bf r}_1$ is the coordinate of 
the struck nucleon denoted as $1$.
We neglect the interference effects 
among different nucleons with which the photon couples.

We decompose the full hamiltonian, $\hat{H}_{A}$, 
in eq.~(\ref{eq:form2}) as 
\begin{eqnarray}
  \hat{H}_{A} = \hat{K}_{1} + \hat{H}_{A - 1} + \hat{V}_{1, A - 1}. 
\label{eq:form4}
\end{eqnarray}
Here $\hat{K}_1$ is the kinetic-energy operator for the struck nucleon,  
$1$, and $\hat{H}_{A - 1}$ is the hamiltonian for the residual nucleus. 
$\hat{V}_{1, A - 1}$ is the interaction of the struck nucleon 
with other nucleons in the target nucleus, 
which is assumed to be expressed as 
a sum of the two-body interactions, $V_{{\rm NN}}({\bf r})$, 
\begin{eqnarray}
    V_{1, A - 1}({\bf r}_{1}; {\bf r}_{2}, \cdots, {\bf r}_{A}) 
  = \sum_{j = 2}^{A} V_{{\rm NN}}({\bf r}_{1} - {\bf r}_{j}). 
\label{eq:form7}
\end{eqnarray}
For numerical calculations 
we construct the nucleon-nucleon potential, $V_{{\rm NN}}({\bf r})$, from 
the phenomenological scattering amplitude 
describing the observed cross sections. 
The potential is complex in the energy region where we are interested in. 
We will leave the details in the Appendix. 

Introducing the Green's function for the nucleon, $1$, 
interacting with the residual nucleus in the fixed-scatterer approximation 
which is given by 
\begin{eqnarray}
  G(\omega; {\bf r}_{1}, {\bf r'}_{1}; 
            {\bf r}_{2}, \ldots, {\bf r}_{A})
  \equiv \langle {\bf r}_{1} | 
    {1 \over \omega - \hat{K}_{1} - \hat{V}_{1, A - 1} + {\it i} \eta}
    |{\bf r'}_{1} \rangle,  
\label{eq:form5}
\end{eqnarray}
the response function is given as
\begin{eqnarray}
  S(\omega, {\bf q})
  &=& -{1 \over \pi}\; 
      {\rm Im}~ \int d{\bf r}_{1} d{\bf r'}_{1} 
                     d{\bf r}_{2} \cdots d{\bf r}_{A} \;
      \exp \{{\it i} {\bf q} \cdot ({\bf r'}_{1} - {\bf r}_{1}) \} 
\label{eq:form8}
\\
  &{}& ~~\times 
         \Psi_{A}^{\ast}({\bf r}_{1}, {\bf r}_{2}, 
                              \ldots, {\bf r}_{A}) \;
         \Psi_{A}({\bf r'}_{1}, {\bf r}_{2}, 
                        \ldots, {\bf r}_{A})\; 
         G(\omega; {\bf r}_{1}, {\bf r'}_{1}; 
                             {\bf r}_{2}, \ldots, {\bf r}_{A}), 
\nonumber
\end{eqnarray}
where $\Psi({\bf r}_{1}, {\bf r}_{2}, \ldots, {\bf r}_{A})$
is the $A$-body nuclear wave function. 
The fixed-scatterer approximation implies that 
the excitation energy, $E_n$, of the nucleus is neglected, 
and that the nucleon-nucleon interaction is assumed to be local
\cite{Eis}. 
The struck nucleon is considered to become energetic enough to justify 
the fixed-scatterer approximation for the FSI 
due to $V_{\rm NN}({\bf r})$. 

The Green's function is now a one-body Green's function 
with a multi-centered potential, 
and we here introduce the eikonal expression for it 
\cite{Har:PR}, 
\begin{eqnarray}
  &{}& G(\omega; {\bf r}_{1}, {\bf r'}_{1}; {\bf r}_{2}, 
                                    \cdots, {\bf r}_{A})  
\nonumber\\
  &=& -{i \over v}\; \delta({\bf b}_{1} - {\bf b'}_{1}) \;
        \theta(z_{1} - z'_{1}) \;
       \exp \{ {\it i} p (z_{1} - z'_{1}) 
             - {{\it i} \over v}
        \int_{z_{1}^{\prime}}^{z_{1}} dz_{1}^{\prime \prime}\;
        \sum_{j = 2}^{A} V_{{\rm NN}}({\bf r''}_{1} - {\bf r}_{j}) \}
\nonumber\\
  &{}& ~+ ~({\rm backward ~ piece}), 
\label{eq:ope12}
\end{eqnarray}
where $p$ is related to the energy loss, $\omega$, by 
\begin{eqnarray}
  \omega + m_{{\rm N}} \simeq E_{p} 
  = \sqrt{p^{2} + m_{{\rm N}}^{2}},  
\label{eq:form10}
\end{eqnarray}
with $m_{{\rm N}}$ being the nucleon mass, and $v$ $= p/E_{p}$. 
We are free to choose the direction of the $z$-axis at this stage, 
but will take it to be in the direction of ${\bf q}$, 
since the exponential in the integrand of eq.~(\ref{eq:form8}) is 
strongly oscillating in this direction for large-$|{\bf q}|$. 
Here we have also neglected $E_{n}$, 
because it is small compared to $\omega$, 
which is roughly 2 ~[GeV]. 
The vectors, ${\bf r}_{1}$, ${\bf r'}_{1}$, and ${\bf r''}_{1}$, 
are thus decomposed as, 
\begin{eqnarray}
  {\bf r}_{1}   = ({\bf b}_{1},  z_{1}), ~~
  {\bf r'}_{1}  = ({\bf b'}_{1}, z'_{1}), ~~
  {\bf r''}_{1} = ({\bf b}_{1},  z''_{1}). 
\label{eq:ope12a}
\end{eqnarray}
The backward piece in eq.~(\ref{eq:ope12}) becomes important 
only when we discuss the sum rule, 
and will be omitted in the following. 
Then the imaginary part of the Green's function is obtained as  
\begin{eqnarray}
  &{}& {\rm Im}~G(\omega; {\bf r}_{1}, {\bf r'}_{1}; {\bf r}_{2}, 
                  \cdots, {\bf r}_{A})  
  \equiv {1 \over 2 i} \; 
          \{ G(\omega; {\bf r}_{1}, {\bf r'}_{1}; {\bf r}_{2}, 
             \cdots, {\bf r}_{A})  
           - G^{\ast}(\omega; {\bf r'}_{1}, {\bf r}_{1}; {\bf r}_{2}, 
                      \cdots, {\bf r}_{A}) \}
\nonumber\\
  &=& -{1 \over v}\; \delta({\bf b}_{1} - {\bf b'}_{1}) \;
         \exp \{{\it i} p (z_{1} - z'_{1})\} 
\nonumber\\
  &{}&\times 
    [ \theta(z_{1} - z'_{1}) \;
        \exp \{- {{\it i} \over v}
        \int_{z_{1}^{\prime}}^{z_{1}} dz_{1}^{\prime \prime}\;
        \sum_{j = 2}^{A} V_{{\rm NN}}({\bf r''}_{1} - {\bf r}_{j}) \}
\nonumber\\
  &{}&  ~~~~~~  + ~ \theta(z'_{1} - z_{1}) \;
        \exp \{ {{\it i} \over v}
        \int_{z_{1}}^{z_{1}^{\prime}}dz_{1}^{\prime \prime}\;
        \sum_{j = 2}^{A} V_{{\rm NN}}^{\ast}({\bf r''}_{1} - {\bf r}_{j}) \} 
     ]. 
\label{eq:form9}
\end{eqnarray}
Substituting this expression into the r.h.s. of eq.~(\ref{eq:form8}), 
we obtain the response function, $S(\omega, {\bf q})$, 
in the Glauber approximation. 

Although the response function is given in a closed form, 
its evaluation contains a multi-dimensional integral 
over the $A$-body density matrix, 
\begin{eqnarray}
  \rho_{A}({\bf r}_{1}, {\bf r'}_{1}; {\bf r}_{2}, \ldots, {\bf r}_{A})  
  = \Psi_{A}^{\ast}({\bf r}_{1}, {\bf r}_{2}, \ldots, {\bf r}_{A})  
  \Psi_{A}({\bf r'}_{1}, {\bf r}_{2}, \ldots, {\bf r}_{A}),  
\label{eq:form13}
\end{eqnarray}
which is, of course, impossible for nuclear matter ($A \rightarrow \infty$). 
We previously carried out such a multi-dimensional integral 
for a light nucleus, $^{16}$O, 
in a similar expression for the nuclear transparency \cite{Sek:PLB}, 
and examined various approximations, which could be used for heavier nuclei
including nuclear matter.  
A possible approximation is to choose the $A$-body density matrix as
\begin{eqnarray}
  \rho_{A}({\bf r}_{1}, {\bf r'}_{1}; {\bf r}_{2}, \ldots, {\bf r}_{A})  
= \rho({\bf r}_{1}, {\bf r'}_{1}) \;  
  \prod_{j = 2}^{A} \rho({\bf r}_{j}) 
  g(|{\bf r'}_{1} - {\bf r}_{j}|) \;g(|{\bf r}_{1} - {\bf r}_{j}|), 
\label{eq:form14}
\end{eqnarray}
where $\rho({\bf r}, {\bf r'})$ is the one-body density matrix  
with $\rho({\bf r})$ $\equiv \rho({\bf r}, {\bf r'})$, and
the one-body density is normalized to one, 
{\it i.e.}, $\int d{\bf r} \; \rho({\bf r})$ $= 1$.
$g(|{\bf r}_{i} - {\bf r}_{j}|)$ describes both dynamical and statistical 
two-body correlation between the nucleon, $1$, and nucleon, $j$. 
This approximation consists of including only the two-body correlation 
and of neglecting the so-called spectator effect 
\cite{Ben:PLB}, \cite{Nik:PL}, 
which amounts to dropping two-body correlations 
among the nucleons, $2, \ldots, A$, 
and has been found accurate enough when the finite interaction-range 
of $V_{{\rm NN}}({\bf r})$ is taken into account \cite{Sek:PLB}. 
With the above $A$-body density matrix, 
the multi-dimensional integral over 
${\bf r}_{2},$ $\ldots, {\bf r}_{A}$ factorizes, {\it i.e.}, 
\begin{eqnarray}
&{}& \int d{\bf r}_{2} \cdots d{\bf r}_{A} \;  
       \exp \{- {{\it i} \over v}
        \int_{z_{1}^{\prime}}^{z_{1}} dz_{1}^{\prime \prime}\;
        \sum_{j = 2}^{A} V_{{\rm NN}}({\bf r''}_{1} - {\bf r}_{j}) \}
   \times 
  \rho_{A}({\bf r}_{1}, {\bf r'}_{1}; {\bf r}_{2}, \ldots, {\bf r}_{A})  
\nonumber \\
&\simeq&  \rho({\bf r}_{1}, {\bf r'}_{1}) \; 
\nonumber \\
&{}& \times
           \left( \int d{\bf r}_{2} \; \rho({\bf r}_{2}) \; 
                g(|{\bf r'}_{1} - {\bf r}_{2}|) \;
                g(|{\bf r}_{1} - {\bf r}_{2}|) \;
       \exp \{- {{\it i} \over v}
        \int_{z_{1}^{\prime}}^{z_{1}} dz_{1}^{\prime \prime}\;
          V_{{\rm NN}}({\bf r''}_{1} - {\bf r}_{2}) \} \right)^{A - 1} 
\nonumber \\
&=&  \rho({\bf r}_{1}, {\bf r'}_{1}) \times
       [ \int d{\bf r}_{2} \; \rho({\bf r}_{2}) \; 
                g(|{\bf r'}_{1} - {\bf r}_{2}|) \;
                g(|{\bf r}_{1} - {\bf r}_{2}|) \;
\nonumber \\
&{}& ~~ \times
      (  1 - 1 + \exp \{- {{\it i} \over v}
        \int_{z_{1}^{\prime}}^{z_{1}} dz_{1}^{\prime \prime}\;
          V_{{\rm NN}}({\bf r''}_{1} - {\bf r}_{2}) \} ) ]^{A - 1} 
\nonumber \\
&\simeq&  \rho({\bf r}_{1}, {\bf r'}_{1}) \times
       \exp [ -(A - 1) \int d{\bf r}_{2} \; \rho({\bf r}_{2}) \; 
                g(|{\bf r'}_{1} - {\bf r}_{2}|) \;
                g(|{\bf r}_{1} - {\bf r}_{2}|) \;
\nonumber \\
&{}& \times
      (  1 - \exp \{- {{\it i} \over v}
        \int_{z_{1}^{\prime}}^{z_{1}} dz_{1}^{\prime \prime}\;
          V_{{\rm NN}}({\bf r''}_{1} - {\bf r}_{2}) \} ) ], 
\label{eq:form15}
\end{eqnarray}
where we have used the relation between the $A$-body 
and the one-body density matrices requiring
\begin{eqnarray}
            \int d{\bf r}_{2} \; \rho({\bf r}_{2}) \; 
                g(|{\bf r'}_{1} - {\bf r}_{2}|) \;
                g(|{\bf r}_{1} - {\bf r}_{2}|) \simeq 1, 
\label{eq:form16}
\end{eqnarray}
and assumed $A$ to be large. 

Thus, we obtain the expression 
for the response function in the following form, 
\begin{eqnarray}
  S(\omega, {\bf q})
  &=& {1 \over \pi v}\; \int d{\bf b}_{1} dz_{1} dz'_{1}\; 
        \rho({\bf r}_{1}, {\bf r'}_{1}) \;
        \exp \{{\it i} (p - q) (z_{1} - z'_{1})\} 
\label{eq:form12}\\
  &{}&~~\times
      \left[ \theta(z_{1} - z'_{1}) \; 
             \exp \{ F({\bf b}_1; z_1, z'_1) \} 
           + \theta(z'_{1} - z_{1}) \;
             \exp \{ F^{\ast}({\bf b}_1; z'_1, z_1) \} 
      \right], 
\nonumber
\end{eqnarray}
where $F({\bf b}_1; z_1, z'_1)$ expresses the effect of the FSI 
(FSI-function), 
\begin{eqnarray}
   F({\bf b}_1; z_1, z'_1) 
   &\equiv& - (A - 1) \int d{\bf r}_{2}\; \rho({\bf r}_{2}) \; 
        g(|{\bf r'}_{1} - {\bf r}_{2}|)\; 
        g(|{\bf r}_{1} - {\bf r}_{2}|)
\nonumber \\
   &{}& \times
        \left( 1 - \exp \{ -{i \over v}\; 
           \int_{z_{1}^{\prime}}^{z_{1}} dz_{1}^{\prime \prime}\;
                V_{{\rm NN}}({\bf r''}_{1} - {\bf r}_{2}) \} \right), 
\label{eq:form17}
\end{eqnarray}
and ${\bf b'}_1$ $= {\bf b}_1$ due to the eikonal approximation, 
eq.~(\ref{eq:ope12}).
Note that we have not used the expansion in terms of 
the nucleon-nucleon potential, 
because the effect of the interaction with each nucleon 
cannot be assumed to be small. 
The expressions, eqs.~(\ref{eq:form12}) and (\ref{eq:form17}),
are our main results of this work. 

One characteristic point of our formulation appears 
in this expression. 
Reflecting the fact that the Green's function for the struck nucleon, 
eq~(\ref{eq:form5}), is off-diagonal with respect to ${\bf r}_1$ 
and ${\bf r'}_1$, 
the correlation function appears as a product of 
$g(|{\bf r'}_{1} - {\bf r}_{2}|)$$g(|{\bf r}_{1} - {\bf r}_{2}|)$.
This feature is missing in 
the formulation based on the optical potential \cite{Ben:PRC}. 
We will discuss the difference in the subsec.~\ref{sec-opt}.

\subsection{Expressions for Nuclear Matter}
\label{sec-expnm}

For later convenience, 
we here derive a simplified expression 
for the response function, eq.~(\ref{eq:form12}), for nuclear matter. 
It becomes 
\begin{eqnarray}
  S(\omega, {\bf q})
  &=& {V \over \pi v}\; \int_{-\infty}^{\infty} dZ\; 
      \widetilde{W}(Z) \;
        \exp \{{\it i} (p - q) Z \} 
\label{eq:nm1}\\
  &{}&~~\times
      [ \theta(Z) \; \exp \{ F(Z) \} 
           + \theta(-Z) \; \exp \{ F^{\ast}(-Z) \} 
      ], 
\nonumber
\end{eqnarray}
where $Z \equiv z_1 - z'_1$, and $V$ is the volume of the system. 
Since we normalize the one-body density to one, 
$\rho({\bf r})$ becomes $\rho_0$ $\equiv 1/V$. 
The FSI-function, eq.~(\ref{eq:form17}), is expressed as 
\begin{eqnarray}
   F(Z) 
   &=& - \rho_{{\rm NM}} \; \int d{\bf r}_{2}\; 
        g(|{\bf r'}_{1} - {\bf r}_{2}|)\; 
        g(|{\bf r}_{1} - {\bf r}_{2}|)
\nonumber \\
   &{}& \times
        \left( 1 - \exp \{ -{i \over v}\; 
           \int_{z_{1}^{\prime}}^{z_{1}} dz_{1}^{\prime \prime}\;
                V_{{\rm NN}}({\bf r''}_{1} - {\bf r}_{2}) \} \right), 
\label{eq:nm1a}
\end{eqnarray}
where $\rho_{{\rm NM}}$ $= A \rho_0$ is the nuclear matter density, and
we use the value, $\rho_{{\rm NM}}$ $= A/V$ $= 0.17 ~[{\rm fm}^{-3}]$. 
Since both the FSI-function and the one-body density matrix become 
functions of the difference of the arguments for the uniform nuclear matter, 
we simply write $F(Z)$ for the FSI-function and $\widetilde{W}(Z)$
for the one-body density matrix in eq.~(\ref{eq:nm1}),
\begin{eqnarray}
  \widetilde{W}(Z) = \widetilde{W}({\bf r} - {\bf r'}) 
          \equiv \rho({\bf r}, {\bf r'}).
\label{eq:nm2}
\end{eqnarray}

Using the fact that $\widetilde{W}(Z)$ is an even function, 
{\it i.e.}, $\widetilde{W}(-Z)$ $= \widetilde{W}(Z)$, 
and introducing a dimensionless function, $\widetilde{w}(Z)$, defined by 
\begin{eqnarray}
  \widetilde{w}(Z) \equiv V \; \widetilde{W}(Z) 
  = \rho({\bf r}, {\bf r'}) / \rho_{0},
\label{eq:nm4}
\end{eqnarray}
we further reduce the expression in the following way, 
\begin{eqnarray}
  S(\omega, {\bf q})
  &=& {V \over \pi v}\; \int_{0}^{\infty} dZ\; 
      \widetilde{W}(Z) \;
\nonumber\\
  &{}&~~\times
        [ \exp \{ {\it i} (p - q) Z + F(Z) \} 
         + \exp \{-{\it i} (p - q) Z + F^{\ast}(Z) \} 
        ]
\nonumber\\
  &=& {2 \over \pi v}\; \int_{0}^{\infty} dZ\; 
      \widetilde{w}(Z) \;
      \exp \{ {\rm Re}~F(Z) \} \;
      \cos \{ (p - q) Z + {\rm Im}~F(Z) \}, 
\label{eq:nm3}
\end{eqnarray}
where we have decomposed 
$F(Z)$ $\equiv {\rm Re}~F(Z) + i {\rm Im}~F(Z)$. 
This is the expression that we use for the numerical calculations 
for the inclusive cross sections in subsec.~\ref{sec-xsec}. 

To compare our formulation with the formulation based on 
the optical potential in subsec.~\ref{sec-opt}. 
it is convenient to define 
the ``convolution" function, $\zeta(Z)$. 
For this purpose we introduce the PWIA response function, 
$S_{0}(\omega, {\bf q})$, which is defined by
\begin{eqnarray}
  S_{0}(\omega, {\bf q})
  &\equiv& {1 \over \pi v}\; \int_{-\infty}^{\infty} dZ\; 
      \widetilde{w}(Z) \;
        \exp \{{\it i} (p - q) Z \}
\nonumber\\
  &=& \int_{-\infty}^{\infty} dZ\; 
      \widetilde{S}_{0}(Z) \; \exp \{{\it i} (p - q) Z \}, 
\label{eq:nm7}
\end{eqnarray}
where $\widetilde{S}_{0}(Z)$ $\equiv \widetilde{w}(Z) / (\pi v)$ 
is the Fourier transform of $S_{0}(\omega, {\bf q})$. 
Note that $\widetilde{S}_{0}(Z)$ is actually a function of $(p - q)$ 
with $p$ related to $\omega$ by eq.~(\ref{eq:form10}), 
due to our approximation of neglecting the residual nucleus 
excitation energies (the fixed-scatterer approximation). 
Using $\widetilde{S}_{0}(Z)$, we can write the response function as 
\begin{eqnarray}
  S(\omega, {\bf q})
  &=&  \int_{-\infty}^{\infty} dZ\;  \widetilde{S}_{0}(Z) \;
        \exp \{{\it i} (p - q) Z \} \; \zeta(Z), 
\label{eq:nm8}
\end{eqnarray}
where $\zeta(Z)$ is our ``convolution" function defined by 
\begin{eqnarray}
  \zeta(Z) &\equiv&  \theta(Z) \; \exp \{ F(Z) \} 
                 + \theta(-Z) \; \exp \{ F^{\ast}(-Z) \}. 
\label{eq:nm9}
\end{eqnarray}
$\zeta(Z)$ includes all the information on the FSI. 
If there is no FSI, then $\zeta(Z)$ $= 1$, and  $S(\omega, {\bf q})$ 
coincides with $S_{0}(\omega, {\bf q})$.

\subsection{Zero-range Approximations}
\label{sec-appr}

In this subsection we discuss the zero-range approximations 
for the nucleon-nucleon potential, $V_{{\rm NN}}({\bf r})$, 
for the FSI-function, eq.~(\ref{eq:form17}), 
which are often employed in the literatures. 
Within our knowledge, the approximations have never been examined 
in the Glauber approach. 
Here we will study them in two stages within our framework, 
and reduce the numerical work greatly. 

The first one is to make the zero-range approximation 
only in the $z$-direction (ZR1). 
This approximation amounts to replacing 
$V_{{\rm NN}}({\bf r})$ by $V_{\perp}({\bf b})\; \delta(z)$. 
The ${\bf b}$-dependence of the last factor on the r.h.s. 
of eq.~(\ref{eq:form17}) factorizes and we obtain 
\begin{eqnarray}
   1 - \exp \{ -{i \over v}\; 
           \int_{z_{1}^{\prime}}^{z_{1}} dz_{1}^{\prime \prime}\;
                V_{{\rm NN}}({\bf r''}_{1} - {\bf r}_{2}) \} 
   &\simeq&  \Gamma({\bf b}_{1} - {\bf b}_{2}) \;
       \theta(z_1 - z_2) \; \theta(z_2 - z_{1}^{\prime}), 
\label{eq:appr1}
\end{eqnarray}
where ${\bf r}_2$ $= ({\bf b}_2, z_2)$. 
The FSI-function, eq.~(\ref{eq:form17}), becomes 
\begin{eqnarray}
   F_{{\rm ZR1}}({\bf b}_1; z_1, z'_1) 
   &=& - (A - 1) \int d{\bf b}_{2}\; 
       \int_{z_{1}^{\prime}}^{z_1} dz_2 \;  
       \rho({\bf b}_{2}, z_{2}) \; 
        g(|{\bf r'}_{1} - {\bf r}_{2}|)\; 
        g(|{\bf r}_{1} - {\bf r}_{2}|)
\nonumber \\
   &{}& \times
        \Gamma({\bf b}_{1} - {\bf b}_{2}), 
\label{eq:appr3}
\end{eqnarray}
where $\Gamma({\bf b})$ is the Fourier transform of the nucleon-nucleon 
scattering amplitude, and 
the definition is shown in eq.~(\ref{eq:pot2}) in the Appendix. 
ZR1 should be a good approximation for the finite-range interaction 
if $|z_1 - z'_1|$ is large, 
because the step function in the $z$-direction 
due to the zero-range approximation will be hard to discriminate from 
the smoothed one over the range of the interaction 
for large $|z_1 - z'_1|$. 

Though the full calculation of the FSI-function for 
the finite-range interaction is probably difficult, 
the behavior of the small $|z_1 - z'_1|$, 
where $|\int_{z_{1}^{\prime}}^{z_{1}} dz_{1}^{\prime \prime}\;$
$V_{{\rm NN}}({\bf r''}_{1} - {\bf r}_{2})|$ is small, 
can be looked into by expanding the l.h.s. of eq.~(\ref{eq:appr1})
in terms of the potential up to the first order. 
We write the expression (\ref{eq:form17}) for 
$F_{{\rm FR}}({\bf b}_1; z_1, z'_1)$ 
\footnote{FR stands for Finite Range.} as
\begin{eqnarray}
   F_{{\rm FR}}({\bf b}_1; z_1, z'_1) 
   &=& - {i \over v} \; (A - 1) 
           \int_{z_{1}^{\prime}}^{z_{1}} dz_{1}^{\prime \prime}\;
      \int d{\bf r}_{2}\; \rho({\bf r}_{2}) \; 
                V_{{\rm NN}}({\bf r''}_{1} - {\bf r}_{2}) 
\nonumber \\
   &{}& \times
        g(|{\bf r'}_{1} - {\bf r}_{2}|)\; 
        g(|{\bf r}_{1} - {\bf r}_{2}|).    
\label{eq:appr2}
\end{eqnarray}
This expression is useful to see 
how $F_{{\rm FR}}({\bf b}_1; z_1, z'_1)$ and that of ZR1 are 
different in the small $|z_1 - z'_1|$ region. 
The numerical results are shown in subsec.~\ref{sec-conv}.

The second one is to make the zero-range approximation 
in all the direction (ZR2). 
One cannot simply replace 
$V_{{\rm NN}}({\bf r})$ by $V_{0} \; \delta({\bf r})$, 
because the appearance of the $\delta$-function in the exponential 
is meaningless, which does not happen for ZR1. 
We have to be careful for determining the expression 
of $\Gamma({\bf b})$ in this approximation. From eq.~(\ref{eq:pot4}), 
$\Gamma({\bf b})$ is written as 
\begin{eqnarray}
    \Gamma({\bf b})
    &=&      {2\pi \over {\it i} |{\bf p}| } \int {d^{2}{\bf q}
                                   \over (2\pi)^{2}} \;
           \exp \{ -{\it i} {\bf q} \cdot {\bf b} \} \; f({\bf q}). 
\label{eq:appr4a}
\end{eqnarray}
The zero-range approximation in the transverse directions 
implies that $f({\bf q})$ is independent of ${\bf q}$. 
Thus eq.~(\ref{eq:appr4a}) becomes 
\begin{eqnarray}
    \Gamma({\bf b})
    &\simeq&   {2\pi \over {\it i} |{\bf p}| } \; f(0) \;
                 \int {d^{2}{\bf q} \over (2\pi)^{2}} \;
            \exp \{ -{\it i} {\bf q} \cdot {\bf b} \}  
\nonumber\\
    &=&   {2\pi \over {\it i} |{\bf p}| } \; f(0) \; 
           \delta^{(2)}({\bf b}).
\label{eq:appr4}
\end{eqnarray}
Substituting this formula into eq.~(\ref{eq:appr3}), we obtain
\begin{eqnarray}
   F_{{\rm ZR2}}({\bf b}_1; z_1, z'_1) 
   &=& - (A - 1) {2\pi \over {\it i} |{\bf p}| } \; f(0) \; 
         \int d{\bf r}_{2}\; \rho({\bf r}_{2}) \; 
        g(|{\bf r'}_{1} - {\bf r}_{2}|)\; 
        g(|{\bf r}_{1} - {\bf r}_{2}|)
\nonumber \\
   &{}& \times
        \delta^{(2)}({\bf b}_{1} - {\bf b}_{2}) \;
       \theta(z_1 - z_2) \; \theta(z_2 - z_{1}^{\prime}) 
\label{eq:appr5}\\
   &=& - (A - 1) {2\pi \over {\it i} |{\bf p}| } \; f(0) \; 
         \int_{z_{1}^{\prime}}^{z_1} dz_{2}\; \rho({\bf b}_{1}, z_2) \; 
        g(|z_{1}^{\prime} - z_{2}|)\; g(|z_{1} - z_{2}|).
\nonumber 
\end{eqnarray}
The two approximations are numerically compared in the form of the 
FSI-function in subsec.~\ref{sec-conv}
and in the inclusive cross section in subsec.~\ref{sec-xsec}.

\newpage
\setcounter{equation}{0}
\section{Comparison with other Formulations}
\label{sec-comp}

In the following two subsections, we compare our formulation 
based on the Glauber approximation 
with the theory of Gersch, Rodriguez, and Smith (GRS theory) 
\cite{GER:72}, \cite{GER:73} 
and a formulation with the optical potential \cite{Ben:PRC}.
For the latter case, we show some numerical estimates 
in the next section.

\subsection{GRS Theory}
\label{sec-grs}

In this subsection, we point out the features of 
the GRS theory \cite{GER:72}, \cite{GER:73}, 
and compare them with those of our formulation 
by using $R(\omega, {\bf q})$ defined in eq.~(\ref{eq:form2a}),  
\begin{eqnarray}
   R(\omega, {\bf q}) 
   \equiv \langle A | \hat{O}^{\dag}({\bf q}) \;
                     {1 \over \omega - \hat{H}_{A} + i \eta} \;
                      \hat{O}({\bf q}) \;
          | A \rangle,    
\label{eq:grs1}
\end{eqnarray}
where $\hat{O}({\bf q})$ 
$(= \exp \{{\it i} {\bf q}\cdot {\bf r}_{1}\})$ 
is the hard scattering operator defined in eq.(\ref{eq:form3}), 
and $\hat{H}_{A} \; | A \rangle$ $= 0$. 
The response function is obtained by taking its imaginary part 
as in eq.~(\ref{eq:form2}). 
We decompose the full hamiltonian, $\hat{H}_{A}$ 
$= \hat{K}_{1} + \hat{H}_{A - 1} + \hat{V}_{1, A - 1}$, 
in the same way as in eq.~(\ref{eq:form4}) of sec.~\ref{sec-glauber}. 
By expanding $R(\omega, {\bf q})$ in $\hat{V}_{1, A - 1}$ 
up to the first order, 
we compare the Glauber theory with the GRS theory. 

First, we consider the expansion of $R(\omega, {\bf q})$ 
in the GRS theory. 
Since the hard scattering operator, $\hat{O}({\bf q})$, 
is the momentum-shift operator, 
it operates on the Green's function in eq.~(\ref{eq:grs1}) 
to shift the momentum. 
The kinetic-energy operator of the struck nucleon, 
$\hat{K}_{1}$ $= \hat{{\bf p}}_1^2  / (2m)$, 
is the only term affected in the hamiltonian, $\hat{H}_{A}$, 
and is shifted as 
\begin{eqnarray}
   \hat{O}^{\dag}({\bf q}) \; \hat{K}_{1} \; \hat{O}({\bf q}) 
 = \hat{K}_{1} + {\hat{{\bf p}}_1 \cdot {\bf q} \over m} 
    + {{\bf q}^2  \over 2 m}. 
\label{eq:grs4}
\end{eqnarray}
Substituting this expression into eq.~(\ref{eq:grs1}), 
we obtain the expansion in terms of $\hat{H}_{A}$, 
\begin{eqnarray}
   R(\omega, {\bf q}) 
   &=& \langle A | {1 \over \omega - \hat{H}_{A} 
                  - \hat{{\bf p}}_1 \cdot {\bf q} / m 
                  - {\bf q}^2 / (2m) } | A \rangle 
\nonumber\\
   &=& \langle A | {1 \over \omega 
                  - \hat{{\bf p}}_1 \cdot {\bf q} / m 
                  - {\bf q}^2 / (2m) } | A \rangle 
\nonumber\\
   &{}& +~ \langle A | {1 \over \omega 
                - \hat{{\bf p}}_1 \cdot {\bf q} / m 
                - {\bf q}^2 / (2m) }\;
           \hat{H}_{A} \;
                  {1 \over \omega 
                - \hat{{\bf p}}_1 \cdot {\bf q} / m 
                - {\bf q}^2 / (2m) }
          | A \rangle + \cdots
\nonumber\\
   &=& R_{0}(\omega, {\bf q}) + R_{1}(\omega, {\bf q}) + \cdots. 
\label{eq:grs5}
\end{eqnarray}
Using $\hat{H}_{A} \; | A \rangle$ $= 0$, 
we can rewrite $R_{1}(\omega, {\bf q})$ as follows:
\begin{eqnarray}
   R_{1}(\omega, {\bf q}) 
   &\equiv&  \langle A | {1 \over \omega 
                - \hat{{\bf p}}_1 \cdot {\bf q} / m 
                - {\bf q}^2 / (2m) }\;
           \hat{H}_{A} \;
                  {1 \over \omega 
                - \hat{{\bf p}}_1 \cdot {\bf q} / m 
                - {\bf q}^2 / (2m) }\;
          | A \rangle 
\label{eq:grs6}
\\
   &=& \langle A | {1 \over \omega 
                  - \hat{{\bf p}}_1 \cdot {\bf q} / m 
                  - {\bf q}^2 / (2m) } \;
          [\hat{H}_{A}, {1 \over \omega 
                  - \hat{{\bf p}}_1 \cdot {\bf q} / m 
                  - {\bf q}^2 / (2m) } ]\;
                  | A \rangle 
\nonumber\\
   &=& \langle A | {1 \over \omega 
                  - \hat{{\bf p}}_1 \cdot {\bf q} / m 
                  - {\bf q}^2 / (2m) } \;
          [\hat{V}_{1, A - 1}, {1 \over \omega 
                  - \hat{{\bf p}}_1 \cdot {\bf q} / m 
                  - {\bf q}^2 / (2m) } ]\;
                  | A \rangle. 
\nonumber
\end{eqnarray}
Only $\hat{V}_{1, A - 1}$ does not commute with 
$\hat{{\bf p}}_1 \cdot {\bf q} / m$. 

$R_{0}(\omega, {\bf q})$ and $R_{1}(\omega, {\bf q})$ 
in eq.~(\ref{eq:grs5}) gives 
$F_{0}({\bf q})$ and $F_{1}({\bf q})$ 
of Rinat and Taragin \cite{RT:NPA}. 
They apply the GRS theory to the $(e, e')$ inclusive reaction 
for the first time. 
In their work, after they obtain the expression of 
$F_{0}({\bf q})$ and $F_{1}({\bf q})$, 
they exponentiate the potential term of $F_{1}({\bf q})$ 
to obtain their final expression. 

Next, we consider the expansion of $R(\omega, {\bf q})$ 
in terms of $\hat{V}_{1, A - 1}$ in the Glauber theory. 
According to the usual perturbation theory we write 
\begin{eqnarray}
   {1 \over \omega - \hat{H}_{A}} 
   &=& {1 \over \omega - \hat{K}_{1} - \hat{H}_{A - 1}} 
\nonumber\\
   &{}& +~ {1 \over \omega - \hat{K}_{1} - \hat{H}_{A - 1}} \;
           \hat{V}_{1, A - 1} \; 
           {1 \over \omega - \hat{K}_{1} - \hat{H}_{A - 1}} 
        + \cdots. 
\label{eq:grs7}
\end{eqnarray}
We apply the fixed-scatterer approximation and 
the eikonal approximation to the above expression. 
By the fixed-scatterer approximation we imply the replacement with 
$\hat{H}_{A - 1}$ to $\bar{E}_{A - 1}$, where
$\bar{E}_{A - 1}$ is the average value of the excitation energy of 
the residual nucleus. 
By the eikonal approximation we imply 
\begin{eqnarray}
 \langle A |  \hat{O}^{\dag}({\bf q}) \; 
   {1 \over \omega - \hat{K}_{1} - \bar{E}_{A - 1} }\; 
   \hat{O}({\bf q})  | A \rangle 
 &=& \langle A |  
{1 \over \omega - \hat{K}_{1} - \hat{{\bf p}}_1 \cdot {\bf q} / m 
- {\bf q}^2 / (2m) - \bar{E}_{A - 1} }  | A \rangle 
\nonumber \\ 
&\simeq& \langle A |  
 {1 \over \omega - \hat{{\bf p}}_1 \cdot {\bf q} / m 
- {\bf q}^2 / (2m) - \bar{E}_{A - 1} }  | A \rangle.
\label{eq:grs8}
\end{eqnarray}
To obtain the second equality, 
we have put $\hat{K}_{1} |A \rangle$ $\simeq 0$, 
because the initial momentum of the struck nucleon is small 
compared to ${\bf q}$. 
Corresponding to the expansion of eq.~(\ref{eq:grs7}), 
$R(\omega, {\bf q})$ of eq.~(\ref{eq:grs1}) is expanded as 
\begin{eqnarray}
   R(\omega, {\bf q}) 
   &=& \bar{R}_{0}(\omega, {\bf q}) + \bar{R}_{1}(\omega, {\bf q}) + \cdots, 
\label{eq:grs9}
\end{eqnarray}
where $\bar{R}_{0}(\omega, {\bf q})$ 
and $\bar{R}_{1}(\omega, {\bf q})$ are written as
\begin{eqnarray}
   \bar{R}_{0}(\omega, {\bf q}) 
   &\equiv&  \langle A | {1 \over \omega 
                - \hat{{\bf p}}_1 \cdot {\bf q} / m 
                - {\bf q}^2 / (2m) - \bar{E}_{A - 1} }
          | A \rangle, 
\label{eq:grs10}\\
   \bar{R}_{1}(\omega, {\bf q}) 
   &\equiv&  \langle A | {1 \over \omega 
                - \hat{{\bf p}}_1 \cdot {\bf q} / m 
                - {\bf q}^2 / (2m) - \bar{E}_{A - 1} }\;
           \hat{V}_{1, A - 1} \;
\nonumber\\
    &{}&  ~~~ \times 
                 {1 \over \omega 
                - \hat{{\bf p}}_1 \cdot {\bf q} / m 
                - {\bf q}^2 / (2m) - \bar{E}_{A - 1} }\;
          | A \rangle. 
\label{eq:grs11}
\end{eqnarray}

Now let us discuss the comparison of the GRS theory with the Glauber theory. 
For up to the first order expansion we can prove the following equation:
\begin{eqnarray}
    R_{0}(\omega, {\bf q}) + R_{1}(\omega, {\bf q}) 
   &=& 
   \bar{R}_{0}(\omega, {\bf q}) + \bar{R}_{1}(\omega, {\bf q}) 
  + \Delta_{1}\bar{R}_{0}(\omega, {\bf q}) 
  + \Delta_{2}\bar{R}_{0}(\omega, {\bf q}), 
\label{eq:grs16}
\end{eqnarray}
where $\Delta_{1}\bar{R}_{0}(\omega, {\bf q})$ 
is the correction to the eikonal approximation defined by
\begin{eqnarray}
  \Delta_{1}\bar{R}_{0}(\omega, {\bf q}) 
   &=&  \langle A | {1 \over \{ \omega 
                - \hat{{\bf p}}_1 \cdot {\bf q} / m 
                - {\bf q}^2 / (2m) \}^2 } \; \hat{K}_{1} 
          | A \rangle, 
\label{eq:grs13}
\end{eqnarray}
and $\Delta_{2}\bar{R}_{0}(\omega, {\bf q})$ 
is the correction to the fixed-scatterer approximation defined by
\begin{eqnarray}
  \Delta_{2}\bar{R}_{0}(\omega, {\bf q}) 
   &=&  \langle A | {1 \over \{ \omega 
                - \hat{{\bf p}}_1 \cdot {\bf q} / m 
                - {\bf q}^2 / (2m) \}^2 } \; 
                (\hat{H}_{A - 1} - \bar{E}_{A - 1})
          | A \rangle.  
\label{eq:grs14}
\end{eqnarray}

The proof of eq.~(\ref{eq:grs16}) is as follows: 
The difference of the zeroth order in $\hat{V}_{1, A - 1}$ is written as
\begin{eqnarray}
   \bar{R}_{0}(\omega, {\bf q}) - R_{0}(\omega, {\bf q}) 
   &=&  \langle A | {1 \over \{ \omega 
                - \hat{{\bf p}}_1 \cdot {\bf q} / m 
                - {\bf q}^2 / (2m) \}^2 } \; \bar{E}_{A - 1} 
          | A \rangle. 
\label{eq:grs12}
\end{eqnarray}
Adding the two correction terms, eqs.~(\ref{eq:grs13})
and (\ref{eq:grs14}), to the zeroth order expression, eq.~(\ref{eq:grs12}), 
we obtain
\begin{eqnarray}
 &{}&  \bar{R}_{0}(\omega, {\bf q}) - R_{0}(\omega, {\bf q}) 
  + \Delta_{1}\bar{R}_{0}(\omega, {\bf q}) 
  + \Delta_{2}\bar{R}_{0}(\omega, {\bf q}) 
\nonumber \\
   &=&  \langle A | {1 \over \{ \omega 
                - \hat{{\bf p}}_1 \cdot {\bf q} / m 
                - {\bf q}^2 / (2m) \}^2 } \; 
             (\hat{K}_{1} + \hat{H}_{A - 1} )
          | A \rangle
\nonumber \\
   &=& - \langle A | {1 \over \{ \omega 
                - \hat{{\bf p}}_1 \cdot {\bf q} / m 
                - {\bf q}^2 / (2m) \}^2 } \; 
              \hat{V}_{1, A - 1} 
          | A \rangle
\nonumber \\
   &=&  R_{1}(\omega, {\bf q}) - \bar{R}_{1}(\omega, {\bf q}), 
\label{eq:grs15}
\end{eqnarray}
where we have used $\hat{H}_{A} \; | A \rangle$ $= 0$. 
This completes the proof of eq.~(\ref{eq:grs16}). 

The relation of the two formulations, eq.~(\ref{eq:grs16}), 
implies that 
the expansion up to this order is exact in the GRS theory, 
while some corrections are needed to equate them in the Glauber theory. 
Thus one can realize that up to the first order in 
$\hat{V}_{1, A - 1}$ the GRS theory includes less approximation 
than the Glauber theory, 
though the corrections should be small in the high energy region. 

If we go beyond the first order, the situation will change. 
In the GRS theory it looks difficult to estimate higher order terms, 
while higher order terms can be summed up in the Glauber theory. 
This is the crucial difference, because the convergence of the series 
is not good from our discussion below eq.~(\ref{eq:form17}) 
in subsec.~\ref{sec-glauber}. 
Since the higher order terms are important, 
we should sum them up. 
Furthermore, the fixed-scatterer approximation 
and the eikonal approximation employed in the Glauber theory 
are established in high-energy regime. 
In this sense, we believe that the Glauber theory is 
superior to the GRS theory for descriptions of 
the high energy reactions.

\subsection{A Formulation with Optical Potential}
\label{sec-opt}

The optical-potential formulation has been used by various authors to
treat the effects of the FSI in inclusive scattering 
(see {\it e.g.},  Refs.~\cite{Ben:PRC}, \cite{Uchi:PLB}, 
\cite{Cio:PLB}, \cite{Hori:PRC}, \cite{Cen:PRC}). 
Here we briefly review the treatment of the FSI 
and of the nuclear correlation 
in the formulation of Benhar, {\it et al.} \cite{Ben:PRC}. 
The main difference from our formulation will be seen to lie 
in the treatment of the nuclear correlation. 

In the optical-potential formulation the effect of FSI is included 
in the convolution form. 
The target nuclear tensor, $W_{\mu \nu}^{A}(q, \omega)$, 
which corresponds to the response function, 
$S(\omega, {\bf q})$, in our approach, is expressed as
\begin{eqnarray}
 W_{\mu \nu}^{A}(q, \omega) 
  &=& \int_{-\infty}^{\infty} d\omega' \; F(\omega - \omega') \;
     W_{\mu \nu, IA}^{A}(q, \omega' - V(q)), 
\label{eq:opt3}
\end{eqnarray}
where $W_{\mu \nu, {\rm IA}}^{A}(q, \omega)$ is 
the target nuclear tensor in the impulse approximation 
which gives the PWIA cross section 
corresponding to $S_{0}(\omega, {\bf q})$ in eq.~(\ref{eq:nm7}). 
$F(\omega)$ is the convolution function expressed as 
\begin{eqnarray}
  F(\omega - \omega') 
  &=& {1 \over 2 \pi} \; \int_{-\infty}^{\infty} dt \;
    e^{{\it i} (\omega - \omega') t} \;  e^{- W(q, t) |t|}
\nonumber \\
  &=& {1 \over \pi} \;{\rm Re}~ \int_{0}^{\infty} dt \;
    e^{{\it i} (\omega - \omega') t} \;  e^{- W(q, t) t}
\nonumber \\
  &=& {1 \over \pi} \; \int_{0}^{\infty} dt \;
    \cos \{(\omega - \omega') t \} \;  e^{- W(q, t) t}.
\label{eq:opt4}
\end{eqnarray}
$V(q) - {\it i} W(q, t)$ is an ``optical potential". 
The $t$-dependence of the imaginary part is introduced to take 
account of the initial-state correlation between the struck nucleon 
and the other nucleons. 
This is thus not the usual optical potential to be used for 
the elastic scatterings and its foundation is not clear. 
The $\omega$-dependence of $V(q)$ and $W(q, t)$ is assumed to be absent. 
If $W(q, t) = 0$, then $F(\omega - \omega')$ $= \delta(\omega - \omega')$. 

The above expression, eq.~(\ref{eq:opt4}), can be directly 
compared with our convolution function, eq.~(\ref{eq:nm9}), 
because both of them contain full information of FSI, 
and appear in the convolution form with the PWIA response function. 
The numerical comparison will be shown in subsec.~\ref{sec-conv}. 

Let us write $z = v t$ instead of $t$ as a variable of $W(q, z)$. 
Using the zero-range approximation for $V_{{\rm NN}}({\bf r})$, 
the authors of Ref.~\cite{Ben:PRC} give 
\begin{eqnarray}
   W(q, z) ={\hbar \over 2}\; \rho_{{\rm NM}} \; 
            v(q)\; \sigma_{{\rm NN}}^{total}(q) \;
            {1 \over z}\; \int_{0}^{z} dz' \; g(z'). 
\label{eq:opt5}
\end{eqnarray}
$g(r)$ is the pair-distribution function, which corresponds 
to our correlation function, defined by \cite{RMP:Ben}
\begin{eqnarray}
  g(r) \equiv \bar{\rho}_{\rm NN}(\rho_{\rm NM}, r) / \rho_{\rm NM}.
\label{eq:opt6}
\end{eqnarray}
$\bar{\rho}_{\rm NN}(\rho_{\rm NM}, r)$ is 
the average two-body density in nuclear matter at density $\rho_{\rm NM}$.
The average two-body density is defined as 
\begin{eqnarray}
  \bar{\rho}_{\rm NN}({\bf r}_{12}) 
 \equiv {1 \over A} \; \int d{\bf R}_{12} \;
  \rho_{\rm NN}({\bf r}_1, {\bf r}_2), 
\label{eq:opt7}
\end{eqnarray}
with ${\bf R}_{12} \equiv ({\bf r}_1 + {\bf r}_2)/2$ and
${\bf r}_{12} \equiv {\bf r}_1 - {\bf r}_2$.
For nuclear matter $\bar{\rho}_{\rm NN}({\bf r}_{12})$ 
is a function of $|{\bf r}_{12}|$,
If there is no correlation, {\it i.e.}, $g(r)$ $= 1$, then 
\begin{eqnarray}
   W(q, z) ={\hbar \over 2}\; \rho_{{\rm NM}} \; 
            v(q)\; \sigma_{{\rm NN}}^{total}(q).
\label{eq:opt8}
\end{eqnarray}

We cannot directly compare the correlation function 
defined in eq.~(\ref{eq:opt6}) with our correlation function 
defined in eq.~(\ref{eq:form14}). 
However judging from how they appear in the term of 
the FSI effect, $W(q, z)$ of eq.~(\ref{eq:opt6}) and 
the FSI-function of eq.~(\ref{eq:form17}), 
it would be appropriate to identify their $g(r)$ with our $g^{2}(r)$.

\newpage
\setcounter{equation}{0}
\section{Numerical Results and Discussion}
\label{sec-nume}

In this section we will show our numerical results 
and the discussions. 

\subsection{``Convolution" Function}
\label{sec-conv}

To calculate the convolution functions defined by eq.~(\ref{eq:nm9}), 
we need a specific form of the correlation function, $g(r)$. 
We take the correlation function of the form 
\begin{eqnarray}
   g(r) = 1 - c_{1}\;e^{-r^{2}/a_{1}^{2}}, 
\label{eq:nume4}
\end{eqnarray}
where $c_{1}$ $= 0.84\;[{\rm fm}]$, and
$a_{1}$ $= 0.7\;[{\rm fm}]$ (Fig.~3), 
which roughly simulates that of Benhar, {\it et al.} 
(Fig.~6 of Ref.~\cite{Ben:PRC}). 
In our figure we plot $g^{2}(r)$ instead of $g(r)$ in order to compare 
it easily with their pair-distribution function, 
because our $g^{2}(r)$ corresponds to their $g(r)$ 
as we discussed at the end of subsec.~\ref{sec-opt}. 
As one can see from those figures, 
the way how our correlation function approaches unity 
is slightly different from theirs. 
The difference in the curvature may cause 
a nontrivial effect on the cross section. 
In this case, we need another Gaussian term in eq.~(\ref{eq:nume4}). 
Since what we would like to see is the difference 
coming from different manipulations for the FSI and the correlation, 
we believe that our choice of the correlation function 
causes no big problem. 

One more thing which we should comment on is the normalization 
of the wave function. 
For the case of a finite nucleus, we should be very careful 
for the normalization including the effect of the nuclear correlation 
\cite{RMP:Ben}. 
Fortunately the correction should be the order of $1/A$, 
which can be neglected for nuclear matter. 

By using eq.~(\ref{eq:nume4}), 
the approximate expressions for 
$F({\bf b}_1; z_1, z'_1)$, eqs.~(\ref{eq:appr3}) and (\ref{eq:appr5}), 
can be calculated explicitly for nuclear matter. 
The expression for ZR1, eq.~(\ref{eq:appr3}), 
where the zero-range approximation is applied 
only in the $z$-direction, becomes
\begin{eqnarray}
F_{{\rm ZR1}}(Z) &\equiv& F_{{\rm ZR1}}({\bf b}_1; z_1, z'_1) 
\label{eq:nm5} \\
   &=& - \rho_{{\rm NM}} \; 
  \int d{\bf b}_{2} \int_{z_{1}^{\prime}}^{z_1} dz_{2}\; 
        g(|{\bf r'}_{1} - {\bf r}_{2}|)\; 
        g(|{\bf r}_{1} - {\bf r}_{2}|) \;
        \Gamma({\bf b}_{1} - {\bf b}_{2}) 
\nonumber \\
   &=& - \rho_{{\rm NM}} \; {f({\bf 0}) \over 2 i |{\bf p}| \gamma}\;
        [ 4 \gamma \pi Z 
        - {c_{1} \pi \over 1 / a_{1}^{2} + 1 / (4 \gamma) }\; 
          a_{1} \; 
          \{ \sqrt{\pi} 
            - \Gamma \left( {1 \over 2}, 
                {Z^2 \over  a_1^2} \right)
          \} 
\nonumber \\
   &{}&  +~ {c_{1}^2 \pi \over 2 / a_{1}^{2} + 1 / (4 \gamma) }\; 
          {a_{1} \over \sqrt{2}} \; 
          \exp \left( - {Z^2 \over 2 a_1^2} \right) \;
          \{ \sqrt{\pi} 
            - \Gamma \left( {1 \over 2}, 
                {Z^2 \over 2 a_1^2} \right)
          \} ],
\nonumber 
\end{eqnarray}
where $Z = z_1 - z'_1$ and $(A - 1) \rho_0$ $\simeq \rho_{{\rm NM}}$.
Here $\Gamma(z, p)$ is the incomplete Gamma function defined by 
\begin{eqnarray}
 \Gamma(z, p) = \int_{p}^{\infty} dt \; e^{-t} t^{z - 1}, ~~
 ({\rm Re}~z > 0).
\label{eq:nume7}
\end{eqnarray}
A useful formula related to the incomplete Gamma function is
\begin{eqnarray}
\int_{a}^{\infty} dx \; e^{-x^2} = 
  {1 \over 2} \; \Gamma \left({1 \over 2}, a^2 \right). 
\label{eq:nume7a}
\end{eqnarray}

The expression for ZR2 where the zero-range approximation is applied 
in the whole direction, eq.~(\ref{eq:appr5}), becomes
\begin{eqnarray}
  F_{{\rm ZR2}} (Z) &\equiv& F_{{\rm ZR2}} ({\bf b}_1; z_1, z'_1) 
\nonumber \\
   &=& - {2\pi \over {\it i} |{\bf p}| } \; f({\bf 0}) \; 
         \rho_{{\rm NM}} \; 
         \int_{z_{1}^{\prime}}^{z_1} dz_{2}\; 
        g(|z_{1}^{\prime} - z_{2}|)\; g(|z_{1} - z_{2}|).
\nonumber \\
   &=& - {2\pi \over {\it i} |{\bf p}| } \; f({\bf 0}) \; 
           \rho_{{\rm NM}} \; 
        [  Z  - c_{1} \;  a_{1} \; 
          \{ \sqrt{\pi} 
            - \Gamma \left( {1 \over 2}, {Z^2 \over a_1^2} \right)  \} 
\nonumber \\
   &{}&  +~ c_{1}^2 \; {a_{1} \over \sqrt{2}} \; 
          \exp \left( - {Z^2 \over 2 a_1^2} \right) \;
          \{ \sqrt{\pi}       
            - \Gamma \left( {1 \over 2}, 
                {Z^2 \over 2 a_1^2} \right)
          \} ],
\label{eq:nm6}
\end{eqnarray}

Using the Gaussian series of the potential, eq.~(\ref{eq:pot7}), 
in the Appendix, 
we write the expression for FR, eq.~(\ref{eq:appr2}), as 
\begin{eqnarray}
  F_{{\rm FR}}(Z) &\equiv&  F_{{\rm FR}}({\bf b}_1; z_1, z'_1) 
\label{eq:nm12} \\ 
  &=& - {i \over v} \; \rho_{{\rm NM}} \; 
           \int_{z_{1}^{\prime}}^{z_{1}} dz_{1}^{\prime \prime}\;
      \int d{\bf r}_{2}\; V_{{\rm NN}}({\bf r''}_{1} - {\bf r}_{2}) \;
        g(|{\bf r'}_{1} - {\bf r}_{2}|)\; 
        g(|{\bf r}_{1} - {\bf r}_{2}|) 
\nonumber \\
   &=& - {i \over v} \; \rho_{{\rm NM}} \; \pi^{3/2} \;
         \sum_{n = 1}^{N_0} V_n \; 
      [ \{ \left( {n \over 4 \gamma} \right)^{-3/2} 
             + c_1^2 \; \left( {2 \over a_1^2} \right)^{-3/2} \;
            \exp \left( - {Z^2 \over 2 a_1^2} \right)     \} Z 
\nonumber \\
   &{}&   -~ c_{1} \; { \{ (n / (4 \gamma)) (1/a_{1}^2) \}^{-1/2} 
     \over n / (4 \gamma) + 1/a_{1}^2 } \;   
          \{ \sqrt{\pi} 
            - \Gamma \left( {1 \over 2}, 
               {  (n / (4 \gamma)) (1/a_{1}^2)  
          \over n / (4 \gamma) + 1/a_{1}^2 } \;   
                Z^2 \right)  \} ].
\nonumber
\end{eqnarray}
Note that this expression is valid only 
for small-$Z$ $(= z_1 - z'_1)$ region. From the discussion 
at the end of the Appendix, 
the expansion should be valid for $|Z|$ $< 0.4$ [fm]
for the parameterization which we use below. 

To clarify the relation of ZR1 with FR we introduce 
one more expression, ZR1'. 
This has no physical implication, 
but is meaningful for obtaining some insight on the relation 
between ZR1 and FR. 
In order to obtain ZR1', we apply the expansion in terms 
of the potential to $\Gamma({\bf b})$ in eq.~(\ref{eq:nm5}), 
\begin{eqnarray}
  \Gamma({\bf b}) &\equiv& 
   1 - \exp \{ -{i \over v} \int_{-\infty}^{\infty} dz' \;
                       V_{{\rm NN}}({\bf b}, z') \}
\nonumber\\
    &\simeq&  {i \over v} \int_{-\infty}^{\infty} dz' \;
                       V_{{\rm NN}}({\bf b}, z'). 
\label{eq:nm14a}
\end{eqnarray}
This approximation is similar to that of FR, 
but the integration range is different. 
Substituting this expression for eq.~(\ref{eq:nm5}), 
we obtain, 
\begin{eqnarray}
F_{{\rm ZR1'}}(Z) &\equiv& F_{{\rm ZR1'}}({\bf b}_1; z_1, z'_1) 
\nonumber \\
   &=& - {i \over v} \;\rho_{{\rm NM}} \; \sqrt{\pi}\;
  \int d{\bf b}_{2} \int_{z_{1}^{\prime}}^{z_1} dz_{2}\; 
        g(|{\bf r'}_{1} - {\bf r}_{2}|)\; 
        g(|{\bf r}_{1} - {\bf r}_{2}|) \;
\nonumber \\
   &{}& \times
        \int_{-\infty}^{\infty} dz''_{1} \;
          V_{{\rm NN}}({\bf r}_{1}^{\prime \prime} - {\bf r}_{2}) 
\\
   &=& - {i \over v} \;\rho_{{\rm NM}} \; \sqrt{\pi} \;
     \sum_{n = 1}^{N_0} V_n \sqrt{{4 \gamma \over n}}  
          [ {4 \gamma \over n}\; \pi Z 
        - {c_{1} \pi \over 1 / a_{1}^{2} + n / (4 \gamma) }\; 
          a_{1} \; 
          \{ \sqrt{\pi} 
            - \Gamma \left( {1 \over 2}, 
                {Z^2 \over a_1^2} \right)
          \} 
\nonumber \\
   &{}&  +~ {c_{1}^2 \pi \over 2 / a_{1}^{2} + n / (4 \gamma) }\; 
          {a_{1} \over \sqrt{2}} \; 
          \exp \left( - {Z^2 \over 2 a_1^2} \right) \;
          \{ \sqrt{\pi} 
            - \Gamma \left( {1 \over 2}, 
                {Z^2 \over 2 a_1^2} \right)
          \} ].
\nonumber 
\label{eq:nm14}
\end{eqnarray}
The first term of this expression is the leading term for large-$Z$. 
This term is the same as that of FR of eq.~(\ref{eq:nm12}). 
Therefore, ZR1' and FR should show the same large-$Z$ behavior. 

The numerical results for the convolution functions are shown 
in Figs.~4 a)-c).
In Figs.~4a) and b) we plot 
\begin{eqnarray}
   {\rm Re}~ \zeta(Z) 
   =  \exp \{ {\rm Re}~F(Z) \} \; \cos \left({\rm Im}~F(Z) \right),
\label{eq:nm13}
\end{eqnarray}
because this form appears in eq.~(\ref{eq:nm3})
for $p$ $= q$, and it carries full information on FSI. 
In Fig.~4a) the results with the nuclear correlation are shown. 
The important finding here is that the slope of FR at $Z$ $= 0$ is 
the same as that of ZR1. 
This implies that the ZR1 can be a good approximation of the finite-range 
interaction case even for small-$Z$. 
Since we consider ZR1 as a good approximation of FR for large-$Z$, 
we expect it to be so for the whole range of $Z$.
ZR2 shows a similar behavior to ZR1 in large-$Z$ region, 
but in small-$Z$ region ZR2 behaves differently from ZR1. 
Thus ZR2 cannot be expected to be generally applicable. 
ZR1' shows the same behavior as FR in the large-$Z$ region, 
and slightly different in the small-$Z$. 
This is what we expected for the small-$Z$ expansion 
and thus justifies our expectation that ZR1 is a good approximation 
for the whole region. 

In Fig.~4b) the results without the nuclear correlation are shown. 
ZR1 and ZR2 degenerate as one sees from eqs.~(\ref{eq:nm5}) and 
(\ref{eq:nm6}), 
and they show different behavior from FR even in small-$Z$. From those results, 
we realize that the nuclear correlation makes 
the zero-range approximations better. 
The nuclear correlation plays an important role in this context. 

In order to compare our formulation with that of the optical potential 
\cite{Ben:PRC}, 
we make the Fourier transform of our convolution function, $\zeta(Z)$, 
because their results were shown as a function of $\omega$, 
\begin{eqnarray}
  \widetilde{\zeta}(\omega) 
  &\equiv&  {1 \over 2 \pi} \; 
    \int_{-\infty}^{\infty} dt \; e^{i \omega t} \; \zeta(t)
\nonumber \\
   &=&  {1 \over 2 \pi} \;\int_{0}^{\infty} dt\;
          \left( \exp \{+i \omega t + F(t) \} 
                + \exp \{-i \omega t + F^{\ast}(-t) \} \right)
\nonumber\\
   &=&  {1 \over \pi} \; \int_{0}^{\infty} dt\;
          \exp \{ {\rm Re}~F(t) \} \;
            \cos \left( \omega t + {\rm Im}~F(t)  \right).
\label{eq:nm11}
\end{eqnarray}
Here we use $v t$ $= Z$. 

The numerical results are shown in Fig.~4c). 
In the figure we plot the Fourier transform of ZR1, ZR2, and ZR2 
without the real part of $V_{{\rm NN}}(r)$.
This figure should be compared with Fig.~4 of Ref.~\cite{Ben:PRC}. 
The characteristic feature of our results of ZR1 and ZR2 is 
the positive slope at $\omega$ $= 0$. 
This behavior comes from the appearance of ${\rm Im}~F(t)$ 
in the argument of cosine of eq.~(\ref{eq:nm11}), 
and is the effect of the real part of $V_{{\rm NN}}(r)$. 
To confirm this statement, we plot 
ZR2 without the real part of $V_{{\rm NN}}(r)$.
The curve turns out to be monotonically decreasing. 

For reference, we plot $\widetilde{\zeta}_{\rm OP}(\omega)$ 
defined by
\begin{eqnarray}
  \widetilde{\zeta}_{\rm OP}(\omega) 
  &\equiv& {1 \over \pi} \; \int_{0}^{\infty} dt \;
    \cos (\omega t ) \;  e^{- W(q, t) t}.
\label{eq:nmxx}
\end{eqnarray}
which corresponds to the case of ${\rm Im}~F(t)$ $= 0$ in eq.~(\ref{eq:nm11}), 
and is the same form as eq.~(\ref{eq:opt4}), {\it i.e.}, 
the correlated Glauber formulation of Ref.~\cite{Ben:PRC}. 
We use eq.~(\ref{eq:opt5}) for $W(q, t)$ with our parameter set, and 
$g(r)$ in eq.~(\ref{eq:opt5}) is replaced our correlation function, 
$\{g(r)\}^2$ of eq.~(\ref{eq:nume4}). 
Therefore, the following expression of $W(q, t)$ is not completely 
the same as that of Ref.~\cite{Ben:PRC}:
\begin{eqnarray}
  W(q, t) 
&\equiv& {1 \over 2}\; \rho_{\rm NM} \; v\; \sigma_{\rm NN}^{total} \;
          {1 \over t}\; \int_{0}^{t} dt' \; \{g(v t)\}^2
\nonumber \\
  &=&  {1 \over 2}\; \rho_{\rm NM} \; v \; \sigma_{\rm NN}^{total} \;
        [  1  
        - c_{1} \;  {a_{1} \over v t} \; 
          \{ \sqrt{\pi} 
            - \Gamma \left( {1 \over 2}, {v^2 t^2 \over a_1^2} \right)  \} 
\nonumber \\
   &{}&  +~ c_{1}^2 \; {a_{1} \over 2 \sqrt{2} v t} \; 
          \{ \sqrt{\pi}       
            - \Gamma \left( {1 \over 2}, 
                {2 v^2 t^2 \over a_1^2} \right)
          \} ].
\label{eq:nmxx2}
\end{eqnarray}
The $q$-dependence of $W(q, t)$ is implicitly included 
in $v$ and $\sigma_{\rm NN}$.
As one can see from Fig.~4c), 
$\widetilde{\zeta}_{\rm OP}(\omega)$ shows a quite similar behavior 
to the result of $\widetilde{\zeta}(\omega)$ of ZR2 without 
Re $V_{\rm NN}$ except for the magnitude.

\subsection{Inclusive Cross Section}
\label{sec-xsec}

We numerically calculate the cross sections of 
the $(e, e')$ inclusive scattering off nuclear matter. 
Since we are interested in the treatment of the FSI and 
the nuclear correlation, 
other information, such as the one-body density matrix, 
is taken as an input. 

We determine the shape of the one-body density matrix, 
$\rho({\bf r}, {\bf r'})$, from numerical results of 
the momentum distribution of a nucleon, $W({\bf k})$. 
The one-body density matrix is related to 
the momentum distribution as \cite{ANT:OX}
\begin{eqnarray}
  \rho({\bf r}, {\bf r'}) 
  \equiv  \rho_{0} \; \int {d{\bf k} \over (2\pi)^{3}}\; 
    e^{-{\it i} {\bf k}\cdot ({\bf r} - {\bf r}')} \; W({\bf k}), 
\label{eq:nume1}
\end{eqnarray}
where $\rho_{0}$ $= 1/V$. 

For the momentum distribution of the nuclear matter in eq.~(\ref{eq:nume1}), 
we use the numerical results of Refs.~\cite{FP:NPA} \cite{PWP:NPA}. 
They use a realistic nuclear force to obtain the results. 
We just use their numbers by $\chi$-square fitting 
in terms of the following series: 
\begin{eqnarray}
  W({\bf k}) &\equiv& (2 \pi)^3 \; n_0(k)
\label{eq:nume2} \\
       &=& (2 \pi)^3 \left(  \theta(k_{F} - k) 
             \{ \alpha 
           + \sum_{j = 1}^{N_{1}} n_{g, j}\;  e^{-j k^{2}/k_{F}^{2}} \}
           + \theta(k - k_{F}) \; 
             \sum_{j = 1}^{N_{2}} n_{e, j}\;  e^{-j k/k_{0}} \right), 
\nonumber
\end{eqnarray}
where $\alpha$ $= 7.156 \times 10^{-2}$, 
$k_{F}$ $= 1.33 ~[{\rm fm}^{-1}]$,  
and $k_{0}$ $= 0.588 ~[{\rm fm}^{-1}]$. 
$N_1$ $= 15$, and $N_2$ $= 10$. 
Since the bare data are given by $n_0(k)$ 
in Refs.~\cite{FP:NPA} \cite{PWP:NPA}, 
we show the relation of $n_0(k)$ with our $W(k)$ in eq.~(\ref{eq:nume2}). 
We impose the condition that the Gaussian series and the exponential 
series are to be connected continuously at $k$ $= k_F$ 
when we calculate the coefficients. 
The normalization is 
\begin{eqnarray}
  \int {d{\bf k} \over (2 \pi)^{3}} \; W({\bf k}) = 1.
\label{eq:nume3}
\end{eqnarray}
The resulting momentum distribution is shown in Fig.~5. 
For comparison, we plot the momentum distribution 
of the Fermi gas, 
$W({\bf k})$ $= W_{0} \; \theta(k_{F} - k)$. 
The tail shows a reflection of the nuclear correlation. 

The numerical result for the one-body density matrix 
obtained from eq.~(\ref{eq:nume1}) is shown in Fig.~6. 
For comparison, we plot that of the Fermi-gas model, 
which takes the form \cite{ANT:OX}, 
\begin{eqnarray}
  \rho({\bf r}, {\bf r'}) 
  =  \rho_{0}\; {3 j_{1}(k_{F} R) \over k_{F} R}, 
\label{eq:nume1a}
\end{eqnarray}
where $R$ $\equiv |{\bf r} - {\bf r'}|$, and
$j_{1}(x)$ $= (\sin x - x \cos x)/ x^{2}$ 
is the spherical Bessel function. 

Putting the above ingredients into eq.~(\ref{eq:nm3}), 
we obtain the electron inclusive cross sections. 
The numerical results are shown in Figs.~7 a)-c). 
For the electron-nucleon cross section we use 
the so-called Rosenbluth cross section \cite{Povh}, 
and average over the proton and the neutron, {\it i.e.}, 
\begin{eqnarray}
   \langle {d\sigma_{{\rm eN}} \over d\Omega} \rangle_{{\rm el, on}}
   \equiv  {1 \over 2} \left( 
      {d\sigma_{{\rm ep}} \over d\Omega}
      + {d\sigma_{{\rm en}} \over d\Omega} \right).
\label{eq:fer13c}
\end{eqnarray}
The comparison of the calculated cross sections, 
ZR1 and ZR2, is shown in Fig.~7a). 
For comparison, we plot the results of PWIA and the case of no correlation. 
The PWIA result directly reflects the momentum distribution 
of a nucleon in the nuclear matter.  
Inclusion of only the FSI, {\it i.e.}, $g(r)$ $= 1$, broadens the cross section, 
but adding the nuclear correlation (ZR1 and ZR2) reduces the broadening. 
That is, the nuclear correlation reduces the effect of the FSI. 
Those are the general features of our results, 
and were also observed in Ref.~\cite{Ben:PRC}. 
By proceeding the zero-range approximation (ZR1 $\rightarrow$ ZR2), 
the cross section comes lower in the low-$\omega$ region. 

In Fig.~7b) we plot the cross sections 
for the cases of ZR1, ZR2, and no correlation
with or without the real part of $V_{\rm NN}(r)$ 
to see the effect of the real part. 
As one can easily see from the figure, 
the effect is small for all cases. From our discussion 
at the end of the previous subsection, 
the real part affects the behavior of the FSI-function 
(see Fig.~4c)), but it does not affect the cross sections. 
We have thus confirmed the validity of neglecting the real part, 
which has often been presumed in the literature. 

In Fig.~7c) we plot our numerical result of the cross section of ZR1 
(times 0.4) with the experimental data. From the discussion 
in subsec.~\ref{sec-conv}, 
the ZR1 includes the major part of the finite-range effect, 
but the cross section of ZR1 still overestimates 
the experimental cross section for low-$\omega$ region, 
although this is a consistent treatment for the inclusive process. 
This would be due to our use of the on-shell $e N$ cross section. 

As we mentioned in the beginning of subsec.~\ref{sec-elem}, 
we should use an off-shell $e N$ cross section for the elementary process 
since the struck nucleon can become far off-the-mass-shell 
in the final state. 
Actually putting a binding effect by shifting the nucleon mass 
to the smaller one shifts the cross section to large-$\omega$. 
Even this simple manipulation can make the agreement of the numerical 
results with the experimental data better, 
but this is only a prescription. 
Since one cannot determine the off-shell cross section 
without a dynamical model describing the internal structure 
of the nucleon and its interaction with the nuclear medium, 
we avoid to get involved in the off-shell problems in this work. 
The problems are interesting and important, 
but we leave the study for our future work. 

In high-$\omega$ region, the experimental data highly exceeds our results, 
because the inelastic channel of the $e N$ cross section 
takes part in the data. 
To describe the cross section in such a high-$\omega$ region 
is out of our scope of this work.

\newpage
\section{Summary and Conclusion}
\label{sec-summary}

We have given here a method to deal with the FSI 
in the high-energy $(e, e')$ inclusive scattering. 
Since we are interested in the quasi-elastic region, 
the relevant degrees of freedom are the nucleonic ones. 
We have applied the Green's function method, 
and used the Glauber approximation for the FSI. 
An advantage of this formulation is that 
the finite-range effect of the nucleon-nucleon interaction and 
the nuclear correlation are included in a systematical way. 

Though the method allows us to express the FSI effect in a closed form, 
the full calculation still requires a large amount of numerical works. 
We have thus examined two kinds of zero-range approximations 
for the nucleon-nucleon interaction, 
which greatly simplify the integrals involved in the closed form. 
The one called ZR1, which uses the zero-range approximation 
only in the longitudinal direction, has been found 
to be accurate enough for the actual calculations. 

The method has then been compared with the treatments of the FSI 
in the other approaches to the response functions 
such as that of Gersch, Rodrigues and Smith, 
which has been applied to the $(e, e')$ process in Ref.~\cite{RT:NPA}, 
and the optical-potential approach used in Ref.~\cite{Ben:PRC}. 

Our principal objective is to propose a unified framework of calculating 
the inclusive $(e, e')$ and the semi-inclusive $(e, e'p)$ responses 
with the particular emphasis on the treatment of the FSI. 
The results of calculation in the present formalism (ZR1) 
for the $(e, e')$ cross section off nuclear matter show strong effects 
of the FSI especially in the low energy transfer ($\omega$) region 
qualitatively similar to those observed in Ref.~\cite{Ben:PRC}. 
The calculated cross section overestimates the experimental one 
by a factor of about 2 in the peak region and by larger factors 
in the low-$\omega$ region. 
This would be partly due to our simple choice of on-shell kinematics 
for the $e N$ elastic cross section. 
We have not discussed the difficult problem of 
constructing the off-shell cross section, 
which would be required if we were to make a serious comparison 
with the observed $(e, e')$ response. 
Instead, we have simply pointed out that, in the inclusive processes, 
the final nucleon is much more off-shell than the initial one. 
Further studies on the in-medium cross section including the off-shell 
kinematics are necessary to draw definitive conclusions on the FSI 
in connection with the color transparency. 

We should critically comment here 
the approximations which we used in this work. 
We have relied on several approximations, 
whose validation needs further and extensive work. 
One approximation is that the total disregard of $E_n$ 
of the residual nucleus, which 
we have used it to obtain the response function, eq.~(\ref{eq:form8}).  
We believe it is good though we have no quantitative basis for it. 
The other approximation is to our choice 
of the form, eq.~(\ref{eq:form14}), for the $A$-body density matrix, 
which is claimed to represent a reasonable approximation 
for the calculation of the nuclear transparency. 
However it might be inadequate for the calculation 
of the low-energy-transfer region of the inclusive cross section. 
These approximations need to be investigated carefully for 
their quantitative validity.

\begin{center}
 {\bf Acknowledgements}
\end{center}

A.K. would like to express his gratitude to 
Profs.~S.C.Pieper and V.R.Pandharipande 
for kindly giving their numerical tables 
of the momentum distributions to him. 
He is supported by the Special Postdoctoral Researcher Program 
at RIKEN. 

This work is supported by 
the Grant-in-Aid for Scientific Research of Monbusho(C-08640355), 
and by the U.S. DOE at CSUN (DE-FG03-87ER40347) and
by the U.S. NSF at Caltech (PHY-9722428 and PHY-9420470).

\newpage 
\begin{center}
   {\bf APPENDIX}
\end{center}
\appendix

\setcounter{equation}{0}
\section{Nucleon-Nucleon Potential}
\label{sec-pot}

In this appendix, we explain how 
we construct the NN potential phenomenologically. 
This is an extension of the method 
introduced in Ref.~\cite{Koh:NP2}. 

We deal with a two-body scattering in free space. 
${\bf p}$ is a  momentum of the incident proton, 
${\bf p'}$ is that of the outgoing proton, 
and the momentum transfer ${\bf q}$ $= {\bf p} - {\bf p'}$. 
Please do not confuse them with the notation 
of the main part of this paper. 
In the following we focus on a kinematic region where
${\bf p} \cdot {\bf q}$ $\simeq 0$. 

The scattering amplitude is defined by 
\begin{eqnarray}
  f({\bf q}) 
  &\equiv&  
    {{\it i} |{\bf p}| \over 2\pi} \int d^{2}b \; \Gamma({\bf b})  \;
           \exp \{{\it i} ({\bf p} - {\bf p'}) \cdot {\bf b} \}   
\label{eq:pot1}\\
  &\simeq&  f(0) \; \exp \{-\gamma {\bf q}^{2}\},  
\label{eq:pot1a}
\end{eqnarray}
where ${\bf b}$ is the impact parameter defined by 
\begin{eqnarray}
  {\bf r} = {\bf b} + z \;{{\bf p} \over |{\bf p}|}, 
\label{eq:pot3}
\end{eqnarray}
and
\begin{eqnarray}
    f(0) = {i + c_0 \over 4 \pi}\; \sigma_{{\rm NN}}^{total} \; |{\bf p}|.  
\label{eq:pot3a}
\end{eqnarray}
The last expression, eq.~(\ref{eq:pot1a}), 
is a phenomenological fit for the forward scattering amplitude. 
$\Gamma({\bf b})$ in eq.~(\ref{eq:pot1}) 
is the scattering amplitude in the coordinate space, 
and it has the following eikonal expression
\begin{eqnarray}
  \Gamma({\bf b}) &\equiv&  1 - \exp \{{\it i} \chi({\bf b})\}.  
\label{eq:pot2}
\end{eqnarray}
$\chi({\bf b})$ is the profile function defined by 
\begin{eqnarray}
    \chi({\bf b})
    &\equiv& -{1 \over v} \int_{-\infty}^{\infty} dz' \;
                       V_{{\rm NN}}({\bf b}, z') 
\label{eq:pot2a}\\
    &=& {1 \over {\it i}} \log(1 - \Gamma({\bf b})). 
\label{eq:pot2b}
\end{eqnarray}
where $V_{{\rm NN}}({\bf r})$ is the nucleon-nucleon potential 
which we would like to obtain here. 

With eq.~(\ref{eq:pot1}),
we analytically obtain the expression of $\Gamma({\bf b})$ 
for the phenomenological fit of $f({\bf q})$, eq.~(\ref{eq:pot1a}), 
by the Fourier transformation, 
\begin{eqnarray}
    \Gamma({\bf b})
    &=&      {2\pi \over {\it i} |{\bf p}| } \int {d^{2}{\bf q}
                                   \over (2\pi)^{2}} \;
            e^{-{\it i} {\bf q} \cdot {\bf b}} f({\bf q}) \nonumber\\
    &\simeq&   \Gamma(0) \; e^{-{\bf b}^{2} / 4\gamma},  
\label{eq:pot4}
\end{eqnarray}
where
\begin{eqnarray}
    \Gamma(0) = {1 \over 2{\it i} |{\bf p}| \gamma}\; f(0).  
\label{eq:pot5}
\end{eqnarray}

Following the original paper of Glauber \cite{Glau:Lec},
we obtain the potential by using the Abel integral equation 
through eq.~(\ref{eq:pot2a}),   
\begin{eqnarray}
  V_{{\rm NN}}(r) 
  &=&     {v \over \pi}{1 \over r}{d \over dr} 
    \int_{r}^{\infty}bdb\; {\chi(b) \over \sqrt{b^{2} - r^{2}}}
\nonumber\\ 
  &=& {v \over \pi}{1 \over r}{d \over dr} 
    \int_{r}^{\infty}bdb\; 
    {\log(1 - \Gamma({\bf b})) \over {\it i}\; \sqrt{b^{2} - r^{2}}}
\nonumber\\ 
  &=& {v \over \pi}{1 \over r}{d \over dr} 
    \int_{0}^{\infty}dy\; {1 \over {\it i}}\; 
    \log(1 - \Gamma(y^{2} + r^{2}))  \nonumber\\ 
  &=& {v \over 2\gamma \pi {\it i}}\; 
    \int_{0}^{\infty}dy\; 
    {\Gamma(0) \; e^{(y^{2} + r^{2})/4\gamma} 
    \over 1 - \Gamma(0) \; e^{(y^{2} + r^{2})/4\gamma}}, 
\label{eq:pot6}
\end{eqnarray}
where $y = \sqrt{b^{2} - r^{2}}$.
We numerically calculate eq.~(\ref{eq:pot6}), 
and obtain the potential. 
We use the following numbers at $|{\bf p}|$ $= 2.0$ [GeV/c] \cite{LB}:
\begin{eqnarray}
  \sigma_{{\rm NN}}^{total} = 43.8\;[{\rm mb}], ~~
  c_0 = -0.14, ~~
  \gamma = 3.37 \times 10^{-6}\;[({\rm MeV})^{-2}], ~~
  v = 0.905.
\label{eq:pot6a}
\end{eqnarray}
The numerical results are shown in Fig.~8. 

For calculations of the response function, 
it is convenient to express the nucleon-nucleon potential 
in terms of the Gaussian series, 
\begin{eqnarray}
  V_{{\rm NN}}^{app}(r) 
  =  \sum_{j = 1}^{N_{0}} V_{j}\; e^{-j r^{2}/4\gamma}.
\label{eq:pot7}
\end{eqnarray}
We determine the coefficients by $\chi^{2}$ fit 
with the number of terms, $N_{0}$ $\simeq 50$.

With the above parameterization, eq.~(\ref{eq:pot6a}), 
the absolute value of the following integral, 
\begin{eqnarray}
 &{}& - {i \over v}  \int_{z'_1}^{z_1} dz_{1}^{\prime \prime} \;
          V_{{\rm NN}}({\bf b}_{1}, z_{1}^{\prime \prime}) 
\\
   &=& - {i \over v}   \sum_{n = 1}^{N_0} V_n \sqrt{{4 \gamma \over n}}  \;
          \exp \{ - {n \over 4 \gamma} \;{\bf b}_1^2 \} \;
         {1 \over 2} \;
       \{ \Gamma \left({1 \over 2}, {n \over 4 \gamma} z_1^{\prime 2} \right)
     - \Gamma \left({1 \over 2}, {n \over 4 \gamma} z_1^2 \right) \}, 
\nonumber 
\label{eq:pot8}
\end{eqnarray}
stays less than unity for $|z_1 - z'_1|$ $< 0.4 \; [{\rm fm}]$ 
with ${\bf b}_1$ $= {\bf 0}$ and $z_1$ $= 0$.

\newpage
\begin{center}
{\Large Figure Captions}
\end{center}
\mbox{}\\
{\large Figure 1}\\
{\normalsize Kinematic notations in the $(e, e')$ inclusive reaction. } \\
\mbox{}\\
{\large Figure 2}\\
{\normalsize Comparison of the off-shellness of the nucleon 
in the medium. 
$\delta m^2 / m_{\rm N}^2$ is plotted as a function of $E / m_{\rm N}$. 
The solid line is the case (I), where 
$V$ $= 300\;[{\rm MeV}]$, $S$ $= -350\;[{\rm MeV}]$. 
The dashed line is the case (II), where 
$V$ $= 0\;[{\rm MeV}]$, $S$ $= -50\;[{\rm MeV}]$. 
The dash-dotted line is the case (III), where
$V$ $= -50\;[{\rm MeV}]$, $S$ $= 0\;[{\rm MeV}]$. 
The sum, $S + V$, is kept to be the same. 
}\\
\mbox{}\\
{\large Figure 3}\\
{\normalsize Nuclear correlation function, $g^{2}(r)$, 
of eq.~(\ref{eq:nume4}) as a function of $r$. 
}\\
\mbox{}\\
{\large Figure 4 a)}\\
{\normalsize Comparison of the two zero-range approximations 
for the nuclear matter with the nuclear correlation. 
Real part of our ``convolution" function, 
$\zeta(Z)$, of eq.~(\ref{eq:nm9}) is plotted as a function of 
$Z$ $(= z_1 - z'_1)$. 
The solid curve is ZR1, eq.~(\ref{eq:nm5}), and
the dashed curve is ZR2, eq.~(\ref{eq:nm6}). 
For reference, we plot ZR1', eq.~(\ref{eq:nm14}), 
by the dot-dashed curve, and 
FR, eq.~(\ref{eq:nm12}), for a limited range of $Z$, 
by the dotted curve. 
}\\
\mbox{}\\
{\large Figure 4 b)}\\
{\normalsize Comparison of the two zero-range approximations 
for the nuclear matter without the nuclear correlation. 
The two are indistinguishable in the figure.
The meaning of the curves is the same as Fig.4 a). 
}\\
\mbox{}\\
{\large Figure 4 c)}\\
{\normalsize Comparison of the two zero-range approximations 
for the nuclear matter with the nuclear correlation. 
Fourier transform of our ``convolution" function, 
$\widetilde{\zeta}(Z)$, of eq.~(\ref{eq:nm11}) is plotted as a function of 
$\omega$. 
The solid curve is ZR1, eq.~(\ref{eq:nm5}), and
the dashed curve is ZR2, eq.~(\ref{eq:nm6}). 
For reference, we plot ZR2 without the real part of 
$V_{\rm NN}(r)$ by the dot-dashed curve, and 
the one based on the optical potential, 
$\widetilde{\zeta}_{\rm OP}(\omega)$ of  eq.~(\ref{eq:nmxx}), 
by the dotted curve. 
}\\
\mbox{}\\
{\large Figure 5}\\
{\normalsize 
Momentum distributions of a nucleon in the nuclear matter 
as a function of $k \; [{\rm fm}^{-1}]$. 
$W(k)$ $\equiv (2 \pi)^3 \; n_0(k)$. 
The solid curve is the result of the fitting, 
eq.~(\protect\ref{eq:nume2}), 
and the dotted line is the case of the Fermi gas. 
The crosses are the data points from 
Refs.~\protect\cite{FP:NPA}, \protect\cite{PWP:NPA}. 
}\\
\mbox{}\\
{\large Figure 6}\\
{\normalsize One-body density matrix of the nuclear matter 
as a function of $|{\bf r} - {\bf r}'|$. 
The solid curve is the case including the nuclear correlation. 
The dashed curve is that of the Fermi gas. 
}\\
\mbox{}\\
{\large Figure 7 a)}\\
{\normalsize 
The cross sections of the inclusive scattering 
off nuclear matter. 
The dotted curve is the case of PWIA. 
The dash-dotted curve is the case of only the FSI 
(no correlation effect). 
The solid curve and the dashed curve are the full calculation 
including both the FSI and the nuclear correlation 
with zero-range approximations. 
The solid curve is ZR1, the zero-range approximation in 
$z$-direction only.
The dashed curve is ZR2, the zero-range approximation in 
the whole direction.
}\\
\mbox{}\\
{\large Figure 7 b)}\\
{\normalsize 
Comparison of the cross sections 
for the cases of ZR1, ZR2, and no correlation 
with (the solid curve) or without 
the real part of $V_{\rm NN}(r)$ (the dashed curve). 
}\\
\mbox{}\\
{\large Figure 7 c)}\\
{\normalsize 
Comparison of the numerical result of the cross section 
and the experimental data \protect\cite{Day:PRC}. 
The solid curve is the case of ZR1 $\times 0.4$.
}\\
\mbox{}\\
{\large Figure 8}\\
{\normalsize Nucleon-nucleon potential, eq.~(\ref{eq:pot6}), 
as a function of $r$ for the case of $|{\bf p}|$ $= 2.0$ [GeV/c], and 
$\gamma$ $= 3.37 \;[({\rm GeV/c})^{-2}]$. 
The dashed curve is the real part, and the solid curve is the 
imaginary part. 
}\\
\mbox{}\\

\end{document}